\documentclass[nocompress]{spie}  %>>> to avoid compression of citations

\usepackage{amsmath,amsfonts,amssymb}
\usepackage{graphicx}
\usepackage[colorlinks=true, allcolors=blue]{hyperref}

\usepackage{subcaption}

\newcommand{\Ls}{\mathcal{L}}
\def\sP{{\mathbb{P}}}

\DeclareMathOperator*{\argmax}{arg\,max}

\title{On the benefits of robust models in modulation recognition}

\author[a]{Javier Maroto}
\author[b]{Gérôme Bovet}
\author[a]{Pascal Frossard}
\affil[a]{EPFL, Switzerland}
\affil[b]{Armasuisse, Switzerland}

\authorinfo{Further author information: \\J.M.: E-mail: javier.marotomorales@epfl.ch}

\begin{document} 
\maketitle

\begin{abstract}
Given the rapid changes in telecommunication systems and their higher dependence on artificial intelligence, it is increasingly important to have models that can perform well under different, possibly adverse, conditions. Deep Neural Networks (DNNs) using convolutional layers are state-of-the-art in many tasks in communications. However, in other domains, like image classification, DNNs have been shown to be vulnerable to adversarial perturbations, which consist of imperceptible crafted noise that when added to the data fools the model into misclassification. This puts into question the security of DNNs in communication tasks, and in particular in modulation recognition. We propose a novel framework to test the robustness of current state-of-the-art models where the adversarial perturbation strength is dependent on the signal strength and measured with the ``signal to perturbation ratio" (SPR). We show that current state-of-the-art models are susceptible to these perturbations. In contrast to current research on the topic of image classification, modulation recognition allows us to have easily accessible insights on the usefulness of the features learned by DNNs by looking at the constellation space. When analyzing these vulnerable models we found that adversarial perturbations do not shift the symbols towards the nearest classes in constellation space. This shows that DNNs do not base their decisions on signal statistics that are important for the Bayes-optimal modulation recognition model, but spurious correlations in the training data. Our feature analysis and proposed framework can help in the task of finding better models for communication systems.
\end{abstract}


\section{Introduction}
\label{sec:intro}

% Classification modulation problem. What and why is important.
With the emergence of deep learning \cite{goodfellow2016deep} and its success across multiple fields of research in the last decade, it was just a question of time that it also reached wireless communication systems. Compared with the previous state-of-the-art approaches, which are mainly based on feature extraction from the signals \cite{Dobre_Abdi_Bar-Ness_Su_2007}, deep neural networks (DNNs) present numerous advantages, like its capability of end-to-end learning that both simplifies the model architecture and improves performance. DNNs have been successful in multiple wireless communication areas like wireless resource allocation \cite{Sun_Chen_Shi_Hong_Fu_Sidiropoulos_2017} anomaly detection \cite{Chalapathy_Chawla_2019}, or modulation recognition, which is the main focus of this work.
Modulation recognition plays a very important role in civilian and military applications, and ranges from detecting daily radio stations and managing spectrum resources to eavesdropping and interfering with enemy radio communications.

% Existence of adversarial perturbations. Importance and references to main works.
While DNNs have improved the overall performance and simplified the traditional modulation recognition systems, recent studies on DNNs in other application domains like computer vision have highlighted security issues in these models \cite{Szegedy_Zaremba_Sutskever_Bruna_Erhan_Goodfellow_Fergus_2014, Moosavi-Dezfooli_Fawzi_Fawzi_Frossard_2017}. Specifically, they have been shown to be vulnerable to adversarial examples, which add a carefully crafted but almost imperceptible perturbation, namely adversarial perturbation, to a real data sample. One would expect that a negligible change in the input would not change the prediction, but that is unfortunately not the case. It implies that DNNs can base their decisions on features that do not seem to be aligned with the target task. Understanding the reasons of such vulnerabilities, how we can robustify DNNs against adversarial perturbations, and, more generally, how we can promote desired properties on the networks, currently forms an active line of research.

% Security concerns.
Several authors \cite{Sadeghi_Larsson_2019,Lin_Zhao_2020,Flowers_Buehrer_Headley_2019} have studied the impact of adversarial perturbations against modulation recognition DNN models and have found that, like the computer vision models, they are vulnerable to attacks. This poses a real threat since DNNs are very popular due to their performance, even though there is a general lack of awareness about their vulnerability in communication systems. For example, adversarial perturbations can provide an efficient way of jamming the wireless communication systems, if the modulation recognition model can be easily attacked.

% Work objective
In this work, we want to raise awareness about the importance of robustness for modulation recognition. We compare traditional feature extraction-based methods with current state-of-the-art DNN models in terms of robustness to adversarial perturbations. We propose two frameworks to test modulation recognition models: a robustness framework where we measure direct performance to adversarial perturbations, which is more reminiscent of computer vision and strives to create models with desirable properties; and a security framework, which represents a more realistic scenario in which the attack does not happen at the receptor but at an intermediate compromised node of the wireless communication system. Additionally, by projecting these perturbations in the constellation diagram space we show that the features learned by the DNNs do not correlate with the features learned by Bayes-optimal maximum likelihood methods. This means that DNNs base their decisions on spurious correlations from the data, making them susceptible to varying channel conditions and other out-of-distribution shifts.

\section{Related work}
\label{sec:related_work}

% Classical approaches. Cumulant preprocessing. 
%   (More details can be found in "Survey of automatic modulation classification techniques").
The problem of being able to recognize which modulation has been applied to a signal is not new. Classical signal processing theory can be used in practice to distinguish between different modulations. There are two main approaches: maximum likelihood and feature recognition methods.
\begin{itemize}
\item \textit{Maximum likelihood methods} \cite{Chung-Yu_Huan_Polydoros_1995,Dobre_Abdi_Bar-Ness_Su_2007,Hameed_Dobre_Popescu_2009} approximate the likelihood function of the received signal with all possible modulations and estimate the most likely modulation. It requires prior knowledge of the channel characteristics, which makes its use difficult in the real world, where unexpected channel corruptions like signal offsets and impulse noise cannot be modeled. However, if priors are correct, then the exact computation of the likelihood function is Bayes-optimal, which leads to a perfect modulation classifier. However, in practice computing the exact likelihood function is unfeasible and sub-optimal approximations are used instead. 
\item \textit{Feature recognition methods} \cite{Xie_Li_Wan_2008,zhou2013design,Zhang_Wang_Wu_Tang_2018} are easier to compute and can achieve performance close to maximum likelihood methods. They generate features such as Fourier transform features and high order cumulants (HOCs), that are then fed to classical machine learning models like KNNs or SVMs. Good performance can be achieved using HOCs in noisy settings \cite{Zhang_Wang_Wu_Tang_2018}. However, this approach requires choosing carefully what features to generate. It also oversimplifies the data received by the model, and the features extracted can be susceptible to changes in the channel condition.
\end{itemize}

% Recent approaches. Explain and reference.
More recently, deep learning \cite{goodfellow2016deep} has been proposed as a better solution to recognize modulations. In comparison with the above approaches, it has two clear advantages. It has low complexity which, unlike maximum likelihood methods, scales linearly, and can be trained end-to-end. This increases the quantity and quality of features extracted from the data compared to feature recognition methods. It also allows for extra flexibility, since it does not require any assumption on the channel conditions but learns them directly from the data.

There are mainly two types of Deep Learning models for modulation recognition. On the one hand, inspired by success in speech recognition, some works \cite{Rajendran_Meert_Giustiniano_Lenders_Pollin_2018,Guo_Jiang_Wu_Zhou_2020} propose architectures based on long short-term memory networks (LSTMs) \cite{Hochreiter_Schmidhuber_1997} for modulation recognition. On the other hand, other works \cite{OShea_Corgan_Clancy_2016,West_OShea_2017,Sadeghi_Larsson_2019} employ convolutional neural networks (CNNs) \cite{Krizhevsky_Sutskever_Hinton_2017}, which have been successful in computer vision thanks to their ability to induce translation invariance on the input. The current state-of-the-art in modulation recognition \cite{OShea_Roy_Clancy_2018} uses a model based on the ResNet architecture \cite{Szegedy_Ioffe_Vanhoucke_Alemi_2016}. However, despite their improved performance, there are drawbacks in deep learning approaches compared with the classical methods. In particular, they are black-box models, making them much less interpretable than their classical counterparts, and they are vulnerable to crafted small energy perturbations, which compromise the security of the system against malicious attacks.

% Adversarial perturbations. Explain and reference.
Recently, there have been multiple works in computer vision \cite{Szegedy_Zaremba_Sutskever_Bruna_Erhan_Goodfellow_Fergus_2014, Moosavi-Dezfooli_Fawzi_Fawzi_Frossard_2017} showing that deep neural networks (DNNs) are vulnerable when imperceptible crafted noise is added to the input, formaing what is commonly referred to as adversarial perturbations. This vulnerability is a concern in terms of security \cite{Cao_Xiao_Yang_Fang_Yang_Liu_Li_2019}, but at the same time its study can give us more insights into what features are learned by the classifier. Adversarial perturbations influence discriminative features \cite{engstrom2019adversarial}, since their inclusion greatly affects the model prediction. In computer vision, it has been shown that DNN features are not aligned with human perception, putting into question the relevance of the features learned by the classifier \cite{ilyas2019adversarial}.

%Threats of Adversarial Attacks… (Lin et al, 2020). What they observed, problems with some experiments (do not take into account difference on signal energy, no defence proposed).
Adversarial perturbations have also been shown to fool DNNs trained for the task of modulation recognition in a recent work \cite{Sadeghi_Larsson_2019}. The authors propose to make the power of the perturbation dependent on the signal power. They fix the ratio between the signal and the perturbation power, or signal-to-perturbation ratio (SPR), when studying the effect of the perturbations. They test two different scenarios: white-box attacks, where the attacker has access to the model gradients, and black-box attacks, where only the prediction can be accessed. For the white-box setting, they use the Fast Gradient Sign Method (FGSM) \cite{Goodfellow_Shlens_Szegedy_2015} to craft the perturbations, and found that adversarial perturbations require much less power than additive white gaussian noise (AWGN) to fool the network. Additionally, for the black-box setting they use universal adversarial perturbations \cite{Moosavi-Dezfooli_Fawzi_Fawzi_Frossard_2017} generated from a different DNN. These perturbations are effective even if the signal is shifted in time, making them useful for asynchronous settings. The authors in \cite{Lin_Zhao_2020} craft fixed magnitude perturbations using a variety of white-box attacks, namely FGSM, Projected Gradient Ascent (PGA)\footnote{The PGA algorithm is commonly referred in the adversarial perturbation literature as Projected Gradient Descent (PGD), which is a misnomer given its definition.} \cite{Madry_Makelov_Schmidt_Tsipras_Vladu_2019}, BIM \cite{Kurakin_Goodfellow_Bengio_2017} and MIM \cite{dong2018boosting}. For the black-box setting they test the same attacks on a surrogate model. The authors in \cite{Flowers_Buehrer_Headley_2019} further expand over prior work \cite{Sadeghi_Larsson_2019} using the bit error rate (BER) as a measure of distortion. Additionally, they consider the scenario where the adversarial perturbation is crafted at the transmitter in order to protect the model against eavesdropping. However, in the work they assume perfect knowledge of the DNN used by the eavesdropper, which we think is unrealistic and would fail if there are several eavesdroppers using different modulation recognition models.

\section{Framework}

In this work, we analyze the robustness of state-of-the-art modulation recognition models to adversarial perturbations. We also measure how secure these models would be in a telecommunications system, where a malicious attacker would try to fool the model. To the best of our knowledge, this is the first work that measures both the robustness and security of multiple state-of-the-art models and compare them with classical models.

% Adversarial perturbations
The objective of untargeted adversarial perturbations is to make the model predict the signal class incorrectly. Adversarial perturbations by definition have to be `imperceptible', that is, they are so small that they should not change the true label of the classification task. In modulation recognition we are interested in small L-inf perturbations. To maximize the fooling rate of the model, we define adversarial perturbations as:
\begin{equation}
\label{eq:adv_pert}
    \delta_i^* = \argmax_{\delta_i}\Ls(x_i + \delta_i, y_i, \theta) \qquad \text{s.t.} \qquad \lVert \delta_i \rVert_{\infty} \leq \varepsilon
\end{equation}
where $\delta_i^*$ is the adversarial perturbation, $x_i$ is the clean signal, $y_i$ is the true label, $\lVert \cdot \rVert_{\infty}$ is the infinity-norm, $\theta$ are the model parameters, $\Ls(\cdot)$ is the model loss function, and $\varepsilon$ is the constraint imposed so that the adversarial perturbation is `imperceptible'.

Finding the adversarial perturbation that maximizes the loss is a difficult optimization problem of its own, and it is computationally complex. Iterative algorithms like PGA \cite{Madry_Makelov_Schmidt_Tsipras_Vladu_2019} are expensive since they require backpropagation of the model as many times as the chosen number of iterations, while one-step algorithms like FGSM only requires one backpropagation pass at the cost of adversarial perturbation strength. We write the objectives of the FGSM and PGA algorithms in equations \eqref{eq:fgsm} and \eqref{eq:pga}, respectively:
\begin{align}
    \label{eq:fgsm} \delta_i^* &\approx \varepsilon \text{sign}(\nabla_x \Ls(x_i, y_i, \theta)) \\
    \label{eq:pga} \delta_i^* &\approx \delta_i^{(K)} \qquad \delta_i^{(k+1)} = \sP_{\mathcal{S}_i}(\delta_i^{(k)} + \beta \varepsilon \nabla_x \Ls(x_i + \delta_i^{(k)}, y_i, \theta))
\end{align}
where $\nabla_x$ is the gradient with respect to the input space, $\varepsilon$ is the radius of Lp ball, $\beta$ is the step size of PGA, $K$ is the number of iterations of PGA, and $\sP_{\mathcal{S}_i}$ is the projection operator onto the region $\mathcal{S}_i$.

% 3. Framework 
% Problem description. Robustness and security frameworks.
% (04-23) Spectrum data is logarithmic. Motivation of SPR over epsilon (signal vs fixed sized perturbations).
% (05-07) Importance of filtering noisy signals before training. Adversarial training makes models less accurate when classifying noisy signals.

\begin{figure}
    \centering
    \begin{subfigure}[b]{0.45\linewidth}
        \centering
        \includegraphics[width=\textwidth]{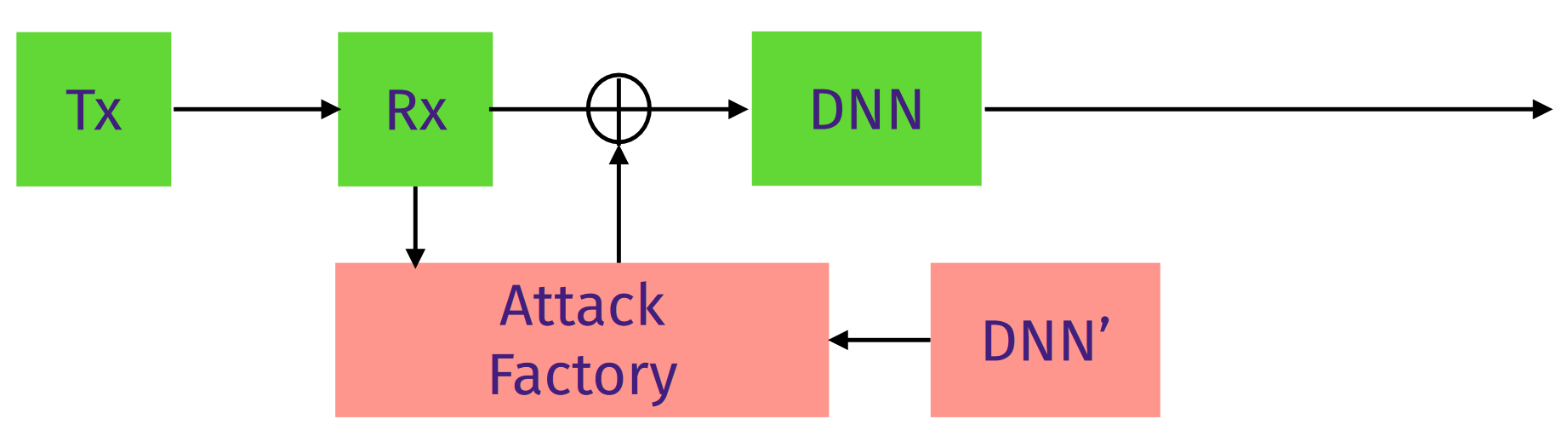}
        \caption{Robustness framework}
        \label{fig:rob_frwk}
    \end{subfigure}
    \begin{subfigure}[b]{0.45\linewidth}
        \centering
        \includegraphics[width=\textwidth]{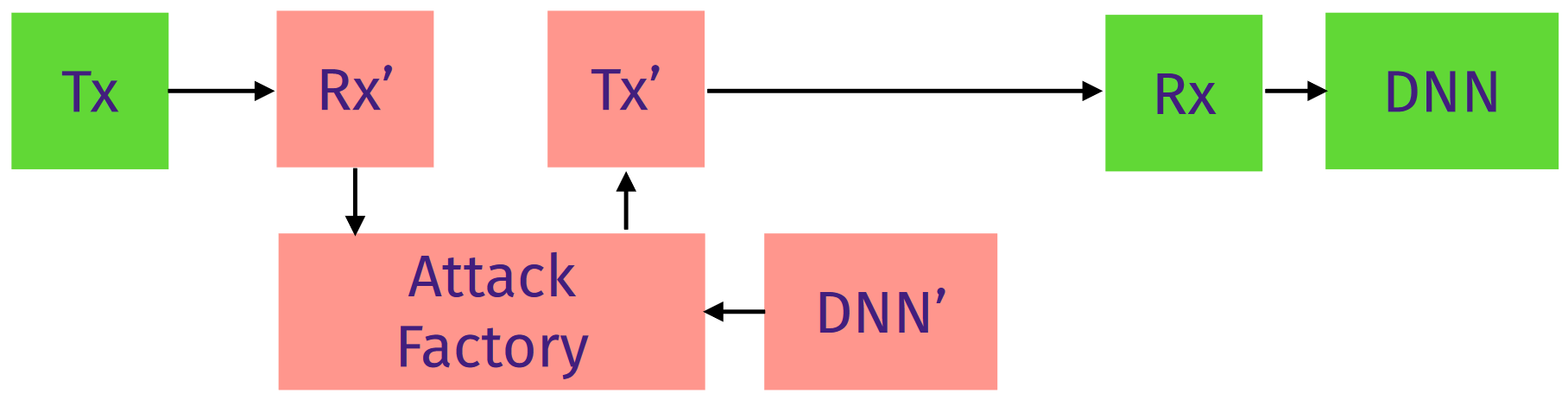}
        \caption{Security framework}
        \label{fig:sec_frwk}
    \end{subfigure}
    \\[2ex]
    \caption{Proposed frameworks for robustness and security. The green and red blocks illustrate the communication system and the attacker, respectively. Tx and Rx are the transmitter and receptor, while DNN is the modulation recognition model. The attacker uses a surrogate model to craft an adversarial perturbation but, since we are analyzing the worst case scenario, we use DNN = DNN' on all our experiments (white-box attack).}
\end{figure}

In computer vision, the perturbation is applied directly before the model. This approach has been used in other fields too. However, in modulation recognition, this is not sufficient to measure how secure the model is against malicious attacks. Different communication effects (e.g. sensor sensitivity) can be modelled as Gaussian noise, and they are added to the received signal. The attacker should not have perfect knowledge of this stochastic process in a realistic scenario. Motivated by that, we propose two frameworks, which have different objectives.
\begin{itemize}
    \item ``Robustness framework" (Figure \ref{fig:rob_frwk}): this is the approach used in computer vision, where the adversarial perturbation is applied before the model. This framework can be used in modulation recognition for analyzing the model, what features it learns, and if those are aligned with the target task. If the model is not susceptible to adversarial perturbations in this framework, it will be secure in realistic security scenarios, where the malicious attacker has less control over the signal that the model receives as input.
    \item ``Security framework" (Figure \ref{fig:sec_frwk}): it represents a scenario adapted to realistic modulation recognition settings. Its purpose is to measure how secure the model is against malicious attacks. The signal is relayed from transmitter to receptor through multiple antennas, and one of them has been compromised. The signal relayed by this antenna is perturbed to fool the modulation classifier at the end of the system. The attacker cannot take into account the posterior stochastic effects, making fooling the network more difficult. For the purpose of this analysis, this scenario can be modeled by adding some noise to the perturbed signal before classifying it.
\end{itemize}

% 4. Experiments 

% (06-05) Bigger gaps between natural and robust accuracy based on dataset size.
% (07-28) Experiment on custom dataset. Big differences in robust accuracy between classes. Intuition of why based on constellation space plots.
\section{Experiments}

\subsection{Datasets}

We now analyze the performance of the classical and state-of-the-art models in the robustness and security frameworks proposed in the previous Section. We use the DeepSig RadioML datasets, concretely the RML2016.10a \cite{OShea_West_2016} and the RML2018.01a \cite{OShea_Roy_Clancy_2018}. They are well-known and widely used datasets for modulation recognition. The RML2016.10a dataset has 220000 IQ signals of 128 time samples each, ranging from -20dB to 18dB signal-to-noise ratio (SNR), where each signal can be one out of 11 different modulations. Although this dataset is the most referenced in the literature, 128 time samples per signal is not enough to extract features from signal statistics (e.g. cumulants), hindering the classical feature recognition methods as well as the deep learning methods to a lesser degree. It also has the issue of the AM-SSB signals having no signal, only AWGN noise. The RML2018.01a dataset was published later to solve most of these problems. It is composed of more than 2.5 million IQ signals of 1024 time samples each, ranging from -20dB to 30dB SNR, where each signal can be one out of 24 different modulations.

We split the RML2016.10a IQ signals into 70\% of training and 30\% of test signals, where 5\% of the training set is used as validation for the hyperparameters of the model, as proposed by \cite{Flowers_Buehrer_Headley_2019}. For the RML2018.01a dataset, we use 1 million signals as training set as proposed in \cite{OShea_Roy_Clancy_2018}, and from the remaining ones we use 5\% for validation and the rest for the test set.

Additionally, we have created a custom dataset (CRML2018) with all the digital modulations of the \linebreak RML2018.01a dataset. Unlike the previous datasets, in this one we have complete knowledge of the symbols transmitted. Furthermore, it is simpler to classify signals since we only corrupt the signal with 20 dB SNR AWGN, which makes this dataset especially interesting to analyze the features learned on the model, as we will explain in the next Section. It has 16 different digital signal modulations. For each of these classes, we have generated 10000 signals of 1024 time samples. We use 70\% of them for training and 30\% for testing, where 5\% of the training ones are used as validation for the hyperparameters of the model.

One important design choice for both our frameworks is the strength of the adversarial perturbation. We advocate for measuring the adversarial strength using the signal-to-perturbation ratio (SPR) instead of using a fixed absolute norm measure like in traditional adversarial computer vision. The main motivation is that IQ signals can differ greatly in power, and since the perceptibility of the adversarial perturbation is related to the signal strength it makes sense to have a relative measure.

\subsection{Performance analysis}

\begin{figure}
    \centering
    \begin{subfigure}[b]{0.45\linewidth}
        \centering
        \includegraphics[width=\textwidth]{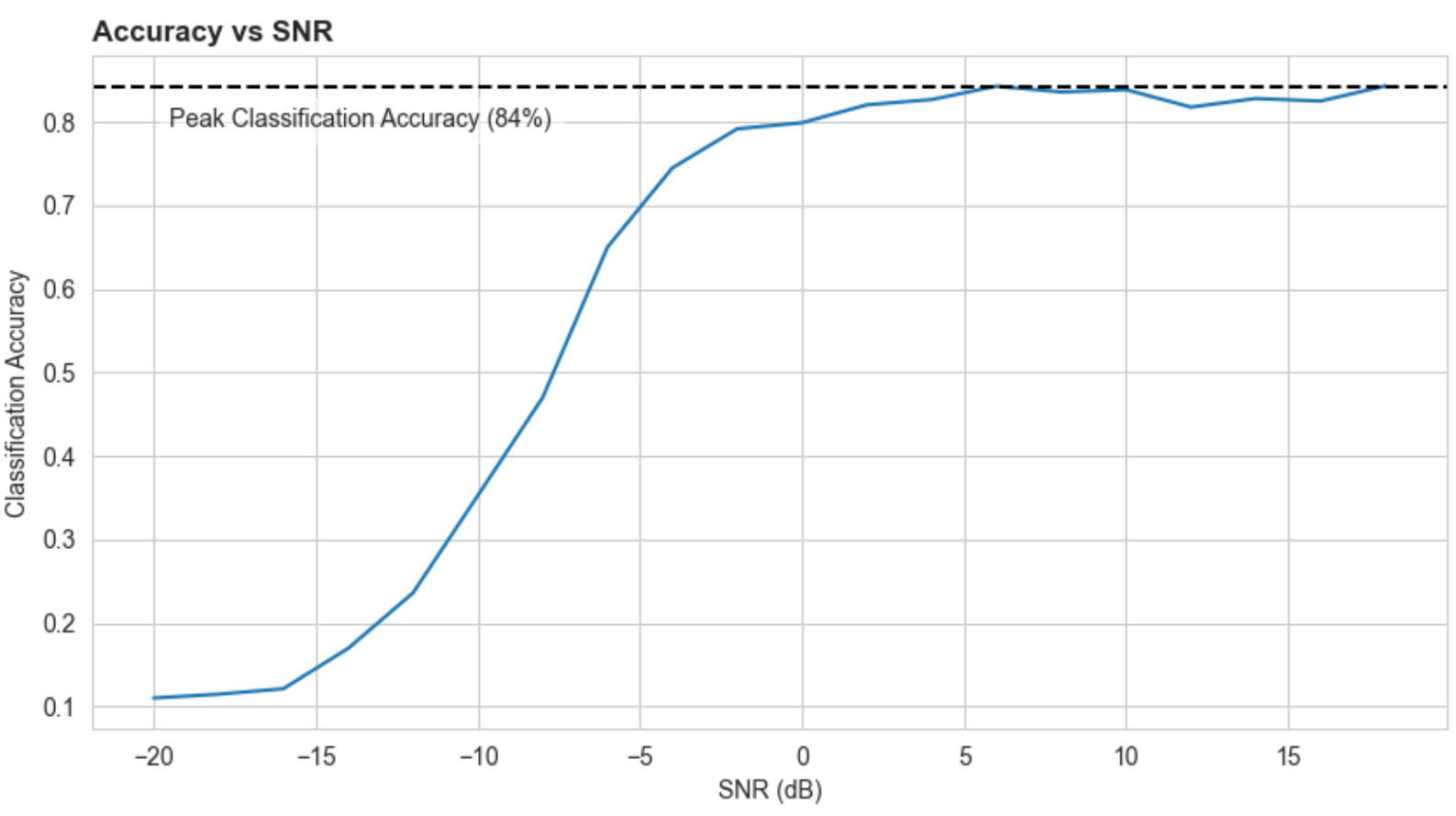}
        \caption{Accuracy on signals from less to higher quality (measured with SNR)}
        \label{fig:snr_sweep}
    \end{subfigure}
    \;
    \begin{subfigure}[b]{0.45\linewidth}
        \centering
        \includegraphics[width=\textwidth]{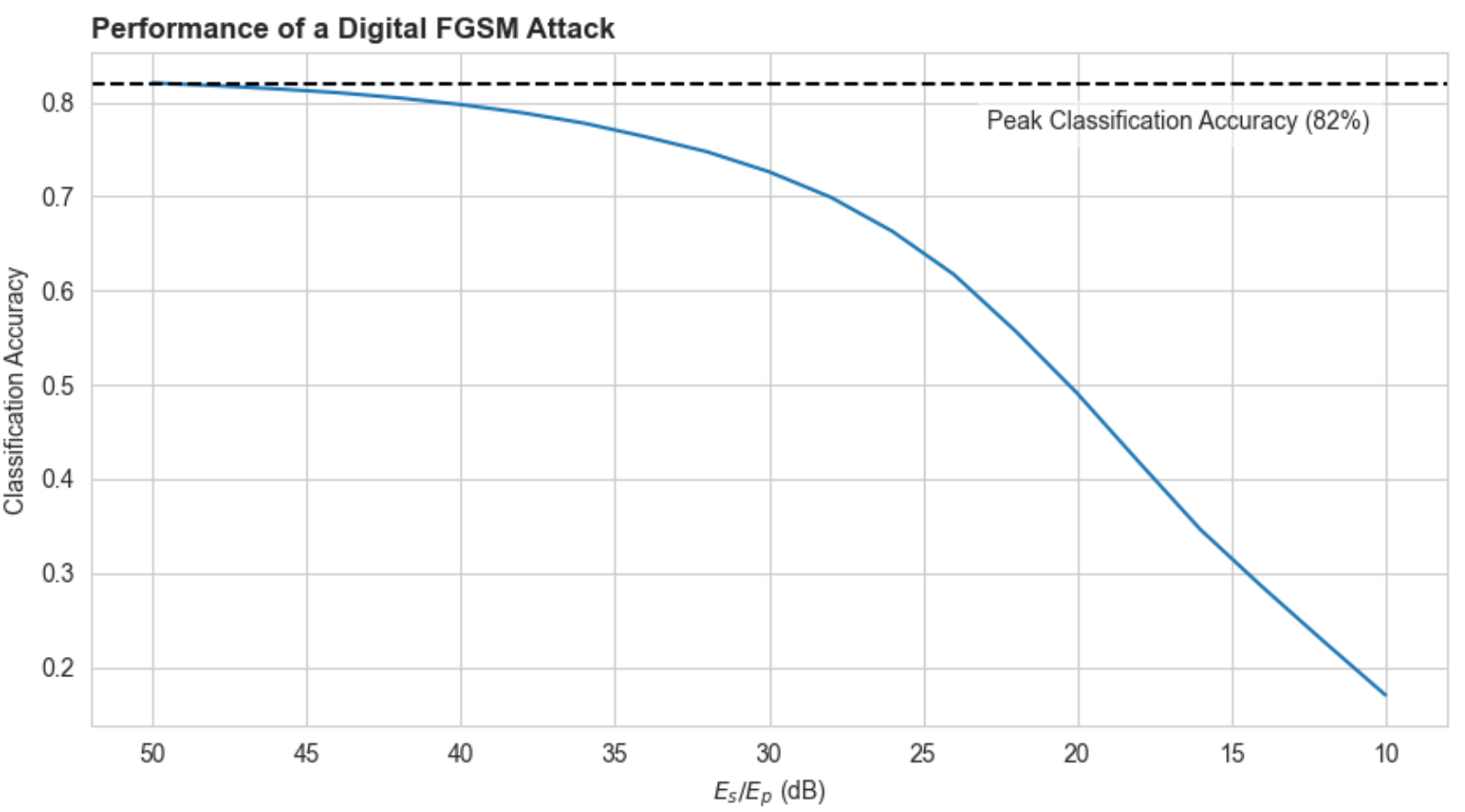}
        \caption{Accuracy on signals with high SNR, but adversarially perturbed with increasing SPR}
        \label{fig:spr_sweep}
    \end{subfigure}
    \\[2ex]
    \caption{VT\_CNN2\_BF performance in the RML2016.10a dataset under different conditions. Much less energy is needed to fool the model with adversarial perturbation compared with AWGN.}
\end{figure}

% (04-23) Reduction of accuracy depending on SPR. Motivate SPR=20dB from graphs.
For our first experiments, we use the robustness framework described before to measure how robust the model is to adversarial perturbations. To choose an effective value of SPR, we test the VT\_CNN2\_BF model \cite{Flowers_Buehrer_Headley_2019}, which has good performance and it is fast to train, on the RML2016.10a dataset under different values of SPR. To minimize the impact of AWGN in the decision we use the cleaner signals from the testing set, which have 18dB SNR. For the attack, we have chosen FGSM because it is faster to compute. Figure \ref{fig:spr_sweep} shows the performance of the model under different levels of perturbation strength. We think using 20dB as the adversarial perturbation SPR is sensible to benchmark the model robustness because not only the accuracy drop is significant but the adversarial perturbation has relatively low power, making it difficult to detect. For comparison, in Figure \ref{fig:snr_sweep} we illustrate the performance of the same model under different SNR conditions and no adversarial perturbation, and we see that it requires a large AWGN energy to cause a similar drop in performance (28dB extra).

\begin{figure}
    \centering
    \begin{subfigure}[b]{0.35\linewidth}
        \centering
        \includegraphics[width=\textwidth]{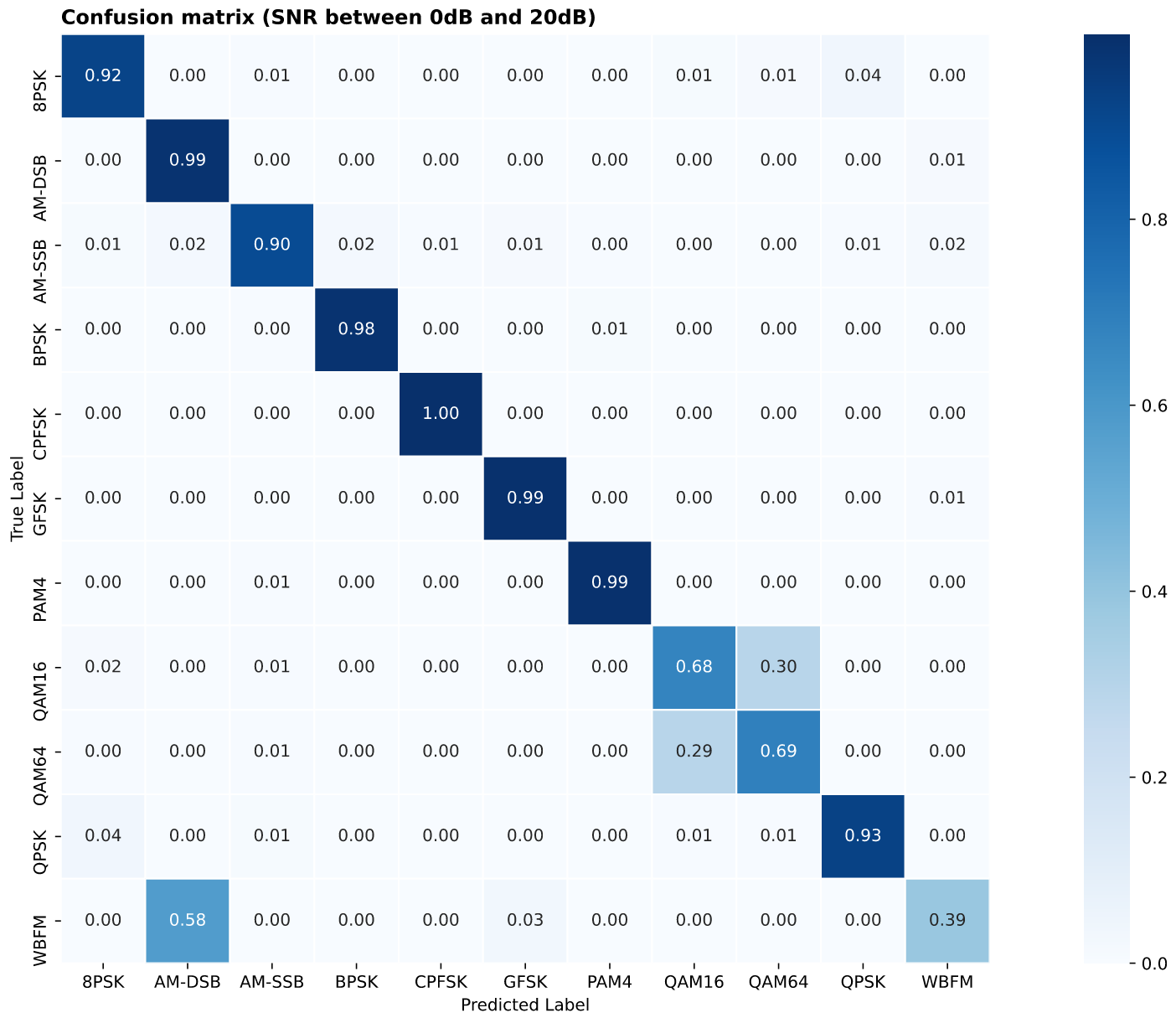}
    \end{subfigure}
    \;
    \begin{subfigure}[b]{0.35\linewidth}
        \centering
        \includegraphics[width=\textwidth]{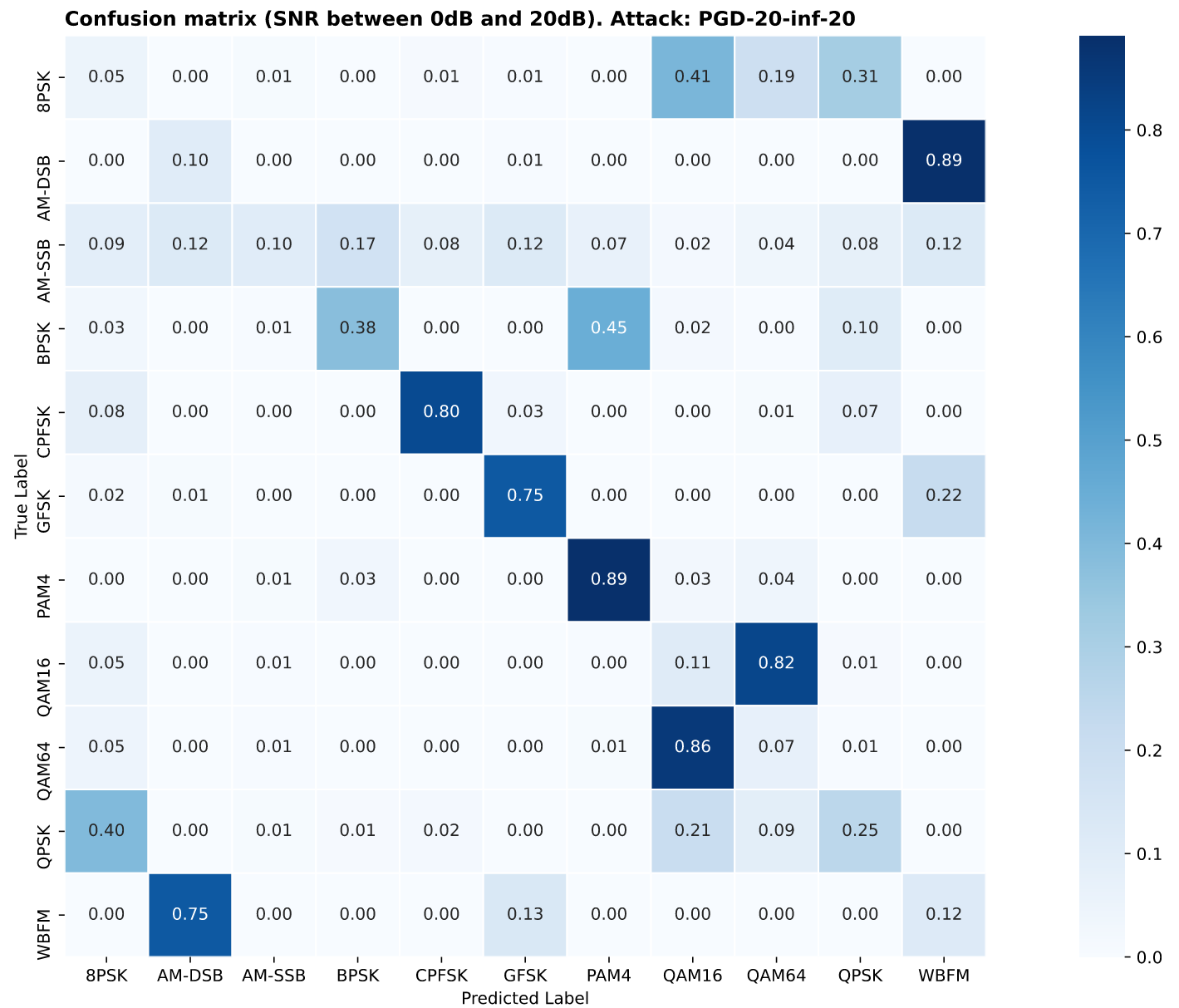}
    \end{subfigure}
    \caption{Confusion matrices of the VT\_CNN2\_BF model before (left) and after (right) adding PGA adversarial perturbations. Results for IQ signals of SNR higher or equal than 0dB}
    \label{fig:cm_std}
\end{figure}
To further motivate the big impact that adversarial perturbations have on the performance of the model, we show in Figure \ref{fig:cm_std} two confusion matrices of the same model when data is or is not perturbed. For the attack, we use PGA of 20dB SPR with 20 iterations of step size 0.125 (relative to the resulting L-inf ball size). We have also tested FGSM, with which we obtain a slightly less significant drop in performance than PGA. This is to be expected since FGSM is a weaker attack \cite{Madry_Makelov_Schmidt_Tsipras_Vladu_2019}. One insight we can obtain from the confusion matrices is the fact that some classes are more robust to attacks than others. This could be especially significant in real scenarios, where some modulations may be preferred in the possible presence of adversarial attacks. Overall, it is quite clear that the model in general is not robust to adversarial perturbations.

\begin{table}[t]
	\centering
	\parbox{.46\linewidth}{
	\begin{tabular}{c|cccc}
		Dataset & Natural & 25dB & 20dB & 15dB \\
		\hline
		RML2016.10a & $0.586$	& $0.444$ & $0.325$	& $0.168$ \\
		RML2018.01a & $0.452$	& $0.099$ & $0.051$	& $0.020$ \\
		CRML2018 & $1.000$	& $0.258$ & $0.048$	& $0.004$ \\
    \end{tabular}
    }
    \hfill
    \parbox{.46\linewidth}{
    \begin{tabular}{c|cccc}
		Dataset & Natural & 25dB & 20dB & 15dB \\
		\hline
		RML2016.10a & $0.586$	& $0.422$ & $0.331$	& $0.198$ \\
		RML2018.01a & $0.452$	& $0.131$ & $0.069$	& $0.033$ \\
		CRML2018 & $1.000$	& $0.273$ & $0.050$	& $0.004$ \\
    \end{tabular}
    }
    \caption{Robustness (left) and security (right) results of the VT\_CNN2\_BF model for different datasets when tested on L-inf perturbations of differing SPR (columns). If ``natural", no perturbation was used. Otherwise, we use PGA with 20 iterations for testing.}
    \label{tab:vt_cnn2_bf_res}
\end{table}

For the following experiments, in order to evaluate both the robustness and the security of the models in realistic scenarios we use our robustness and security frameworks, respectively. To model the security framework, we add 20 dB SNR AWGN to the perturbed signal. We tested the performance of the VT\_CNN2\_BF model against perturbations with varying strength for the RML2016.10a, RML2018.01a, and our CRML2018 dataset. We show the accuracy on the unperturbed signals, and the signals with L-inf PGA perturbations of SPR 15dB, 20dB, and 25dB in Table \ref{tab:vt_cnn2_bf_res}. When comparing the robustness and the security of the model we see almost no difference. We believe the features learned by the model are quite invariant to Gaussian noise, which explains the small variation in performance. The security framework may look more challenging for the attacker, but it is in fact not the case, as we achieve high fooling rates in both frameworks. This result is concerning, as it implies that adversarial perturbations can fool the network no matter the channel conditions.

\begin{table}[!t]
	\centering
	\parbox{.46\linewidth}{
	\begin{tabular}{c|cccc}
		Model & Natural & 25dB & 20dB & 15dB \\
		\hline
		HOC model & $0.977$	& $0.278$ & $0.250$	& $0.000$ \\
		VT\_CNN2\_BF & $1.000$	& $0.500$ & $0.485$	& $0.259$ \\
		ResNet & $1.000$	& $0.393$ & $0.279$	& $0.152$ \\
    \end{tabular}
    }
    \hfill
    \parbox{.46\linewidth}{
    \begin{tabular}{c|cccc}
		Model & Natural & 25dB & 20dB & 15dB \\
		\hline
		HOC model & $0.977$	& $0.273$ & $0.250$	& $0.000$ \\
		VT\_CNN2\_BF & $1.000$	& $0.500$ & $0.486$	& $0.263$ \\
		ResNet & $1.000$	& $0.393$ & $0.395$	& $0.168$ \\
    \end{tabular}
    }
    \caption{Robustness (left) and security (right) results on our custom dataset with only 4 modulations for different models when tested on L-inf perturbations of differing SPR (columns). If ``natural", no perturbation was used. Otherwise, we use PGA with 20 iterations for testing.}
    \label{tab:model_comparison}
\end{table}

Finally, we compare the robustness and security of the fast-to-train VT\_CNN2\_BF model, the current state-of-the-art ResNet model proposed by O'Shea, and a classical feature extraction model based on HOC (with logistic regression being the classifier). Due to the limitations of the latter model, we compare the three of them in a subset of our custom CRML2018 dataset, and we select the four classes that can be classified by the classical feature extraction model (BPSK, QPSK, 16QAM, and 64QAM). The results are shown in Table \ref{tab:model_comparison}. One key result from this experiment is that the better performing ResNet model is more susceptible than the simpler VT\_CNN2\_BF model. This shows that we should not choose the modulation recognition model only based on its accuracy. However, simplicity is not the reason for being more robust. In fact, both the DNN models perform better not only with unperturbed signals but also with perturbed ones, when compared with the cumulants model. This can be explained by the fact that cumulants are computed from signal autocorrelations. While AWGN is non-correlated, adversarial perturbations are highly dependent both on the features learned by the model as well as the input signal itself, which can affect greatly the cumulant values. These results motivate testing the models against adversarial perturbations to ensure that they are safer.

% 5. Analysis 
% (07-12) Model-based properties. Constellation space gives a mathematical probability for each modulation. Easy to visually distinguish clear classes from “gray areas” unlike image classification.
% (07-12) Visualization of perturbations on constellation space (BPSK vs QPSK). Models seem to learn correct directions.
\section{Feature analysis}

In this Section we analyze in more detail the features that the model is learning by observing how aligned the perturbations are with the signal statistics used by maximum likelihood methods. The features learned by the model are a key factor on why they are susceptible to adversarial perturbations, which motivate this analysis. Using the perturbations generated by an attacker to visualize the features learned by the model is not a new concept. Several works in computer vision claim that adversarial perturbations are indicators of the features that the model has learned \cite{Ilyas_Santurkar_Tsipras_Engstrom_Tran_Madry_2019,Engstrom_Ilyas_Santurkar_Tsipras_Tran_Madry_2019,Ortiz-Jimenez_Modas_Moosavi-Dezfooli_Frossard_2020}. The features learned by the model can be classified into two groups: robust features, which are aligned with the true predictors of the task, and non-robust features, which despite generalizing well between train and test hinder the model accuracy in the adversarial setting. The latter ones generally have small norm and adversarial attacks exploit them to fool the network, because their correlation with the label can be easily flipped \cite{Ilyas_Santurkar_Tsipras_Engstrom_Tran_Madry_2019}.

In the context of modulation recognition, this insight is even more useful. Unlike image classification, modulation recognition is a task that can be solved mathematically. Maximum likelihood methods in fact can give us, under some assumptions, an exact probability for a signal to be one particular modulation. While in image classification is not clear which is the shortest path to changing the true class of an image, in modulation recognition it is much more simple. Essentially, a targeted attack should shift each symbol to the closer state in the target modulation. Typically, IQ signals are projected using constellation diagrams \cite{andrew1996tanenbaum} to better visualize how likely a signal belongs to a particular modulation. In this work, we make use of this constellation space to visualize the adversarial perturbations. By visualizing the shifts that the perturbation does on the modulation symbols, we can verify if the model is learning meaningful features. Since we know what should be the optimal way to fool a perfect Bayes-optimal model, we can check how aligned is the perturbation crafted from the trained model, with the optimal one.

\begin{figure}[t]
    \centering
    \begin{subfigure}[b]{0.35\linewidth}
        \centering
        \includegraphics[width=\textwidth]{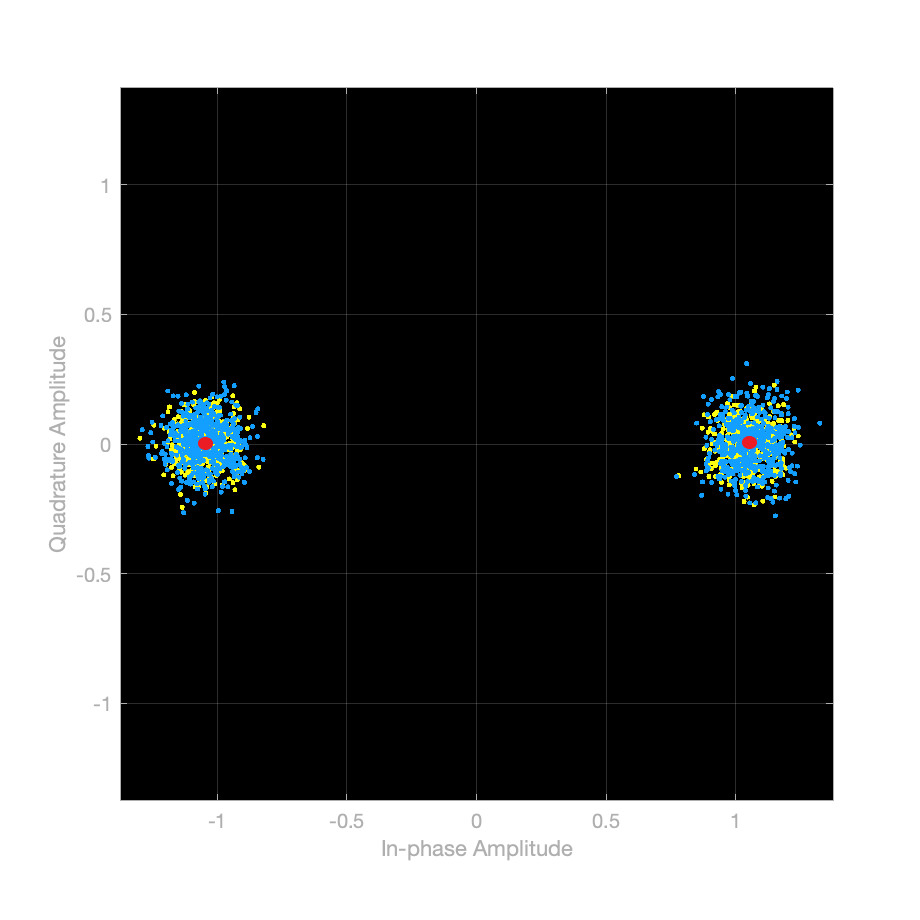}
    \end{subfigure}
    \;
    \begin{subfigure}[b]{0.35\linewidth}
        \centering
        \includegraphics[width=\textwidth]{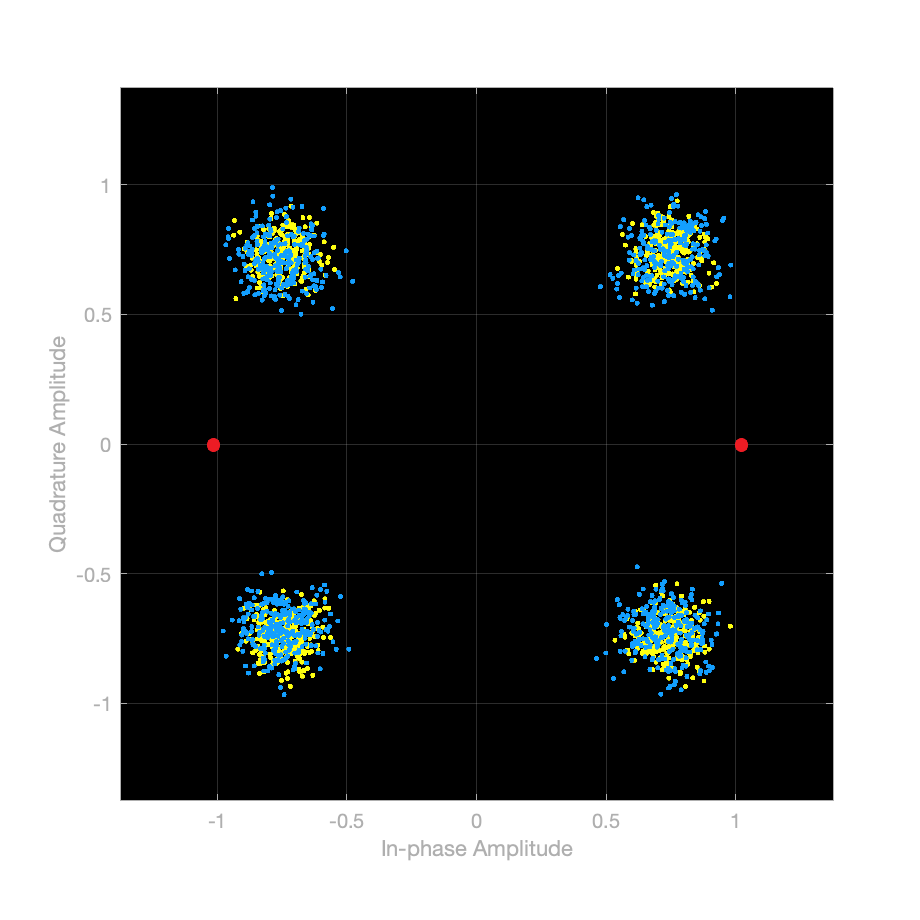}
    \end{subfigure}
    \caption{Constellation diagrams of both original (yellow) and perturbed (blue) samples of a BPSK (left) and QPSK signal (right). We use the VT\_CNN2\_BF model to construct the perturbations. The perturbations are targeted towards maximizing the probability of the signal being BPSK. A Bayes-optimal model would shift the symbols towards the red points, which differs greatly from what we see with the DNN in the plots.}
    \label{fig:cd_mpsk}
\end{figure}

To test our hypothesis that adversarial perturbations should be aligned with the Bayes-optimal perturbation, we trained a VT\_CNN2\_BF model on our CRML2018 dataset and generated a targeted adversarial example towards BPSK using FGSM for one test signal of each class. The objective of targeted adversarial networks is to make the network predict a particular chosen target class. Thus, by using BPSK as a target, we can see how the model thinks we should perturb the signal such that the BPSK class is maximized. Because we only corrupted the signals with Gaussian noise, we know that this perturbation should shift the symbols towards the BPSK valid states. This is what the Bayes-optimal model would do (maximum likelihood). We show in Figure \ref{fig:cd_mpsk} the constellation diagram of each of the signals with both the original and the perturbed samples superposed. While there is a very slight general shift of the symbols towards the BPSK states in the perturbed QPSK signal, most of the symbols are not perturbed in the correct direction. That means that the model did not learn what we know for certain should maximize the BPSK probability of the signal. Thus, if in a very simple scenario the features learned by the model do not seem to be correlated with the task, we can expect the model to be susceptible not only to adversarial perturbations, but possibly to other out-of-distribution shifts.

\section{Conclusion}

In this work, we have proposed two different frameworks to test the vulnerability of DNNs to adversarial perturbations in modulation recognition tasks. The robustness framework is good analysing features learned by the model, while the security framework explores the susceptibility of the model to adversarial perturbations. We showed that, like in other fields, modulation recognition DNNs are susceptible to adversarial perturbations. Moreover, we demostrate that some classical systems like HOC feature extraction models are more susceptible than DNNs. A very helpful insight from this work is that some modulations are less sensitive to perturbations than others, which can be used in the interest of secure communications. We also compare the features learned by DNNs with the Bayes-optimal Maximum Likelihood model, and we see that these features do not seem to be optimal for that task. We encourage future research to test modulation recognition models against adversarial perturbation to ensure not only the safety of the system against attackers, but also make them robust against other out-of-distribution shifts thanks to the model learning better features.

\section{Acknowledgements}

This work has been sponsored by armasuisse Science and Technology with the project ROBIN (project code Aramis 047-22).

% References
\bibliography{references} % bibliography data in report.bib

\begin{thebibliography}{10}

\bibitem{goodfellow2016deep}
Goodfellow, I., Bengio, Y., Courville, A., and Bengio, Y.,  [{\em Deep
  learning}{\nolinebreak\hspace{0.1em}]}, vol.~1, MIT press Cambridge (2016).

\bibitem{Dobre_Abdi_Bar-Ness_Su_2007}
Dobre, O.~A., Abdi, A., Bar-Ness, Y., and Su, W., ``Survey of automatic
  modulation classification techniques: classical approaches and new trends,''
  {\em IET Communications}~{\bf 1},  137–156 (Apr 2007).

\bibitem{Sun_Chen_Shi_Hong_Fu_Sidiropoulos_2017}
Sun, H., Chen, X., Shi, Q., Hong, M., Fu, X., and Sidiropoulos, N.~D.,
  ``Learning to optimize: Training deep neural networks for wireless resource
  management,'' in [{\em 2017 IEEE 18th International Workshop on Signal
  Processing Advances in Wireless Communications
  (SPAWC)}{\nolinebreak\hspace{0.1em}]},   1–6 (Jul 2017).

\bibitem{Chalapathy_Chawla_2019}
Chalapathy, R. and Chawla, S., ``Deep learning for anomaly detection: A
  survey,'' {\em arXiv:1901.03407 [cs, stat]}  (Jan 2019).
\newblock arXiv: 1901.03407.

\bibitem{Szegedy_Zaremba_Sutskever_Bruna_Erhan_Goodfellow_Fergus_2014}
Szegedy, C., Zaremba, W., Sutskever, I., Bruna, J., Erhan, D., Goodfellow, I.,
  and Fergus, R., ``Intriguing properties of neural networks,'' {\em
  arXiv:1312.6199 [cs]}  (Feb 2014).
\newblock arXiv: 1312.6199.

\bibitem{Moosavi-Dezfooli_Fawzi_Fawzi_Frossard_2017}
Moosavi-Dezfooli, S.-M., Fawzi, A., Fawzi, O., and Frossard, P., ``Universal
  adversarial perturbations,'' {\em arXiv:1610.08401 [cs, stat]}  (Mar 2017).
\newblock arXiv: 1610.08401.

\bibitem{Sadeghi_Larsson_2019}
Sadeghi, M. and Larsson, E.~G., ``Adversarial attacks on deep-learning based
  radio signal classification,'' {\em IEEE Wireless Communications
  Letters}~{\bf 8},  213–216 (Feb 2019).

\bibitem{Lin_Zhao_2020}
{Lin}, Y., {Zhao}, H., {Tu}, Y., {Mao}, S., and {Dou}, Z., ``Threats of
  adversarial attacks in dnn-based modulation recognition,'' in [{\em IEEE
  INFOCOM 2020 - IEEE Conference on Computer
  Communications}{\nolinebreak\hspace{0.1em}]},   2469--2478 (2020).

\bibitem{Flowers_Buehrer_Headley_2019}
Flowers, B., Buehrer, R.~M., and Headley, W.~C., ``Evaluating adversarial
  evasion attacks in the context of wireless communications,'' {\em
  arXiv:1903.01563 [cs, eess, stat]}  (Mar 2019).
\newblock arXiv: 1903.01563.

\bibitem{Chung-Yu_Huan_Polydoros_1995}
Huan, C.-Y. and Polydoros, A., ``Likelihood methods for mpsk modulation
  classification,'' {\em IEEE Transactions on Communications}~{\bf 43},
  1493–1504 (Feb 1995).

\bibitem{Hameed_Dobre_Popescu_2009}
Hameed, F., Dobre, O.~A., and Popescu, D.~C., ``On the likelihood-based
  approach to modulation classification,'' {\em IEEE Transactions on Wireless
  Communications}~{\bf 8},  5884–5892 (Dec 2009).

\bibitem{Xie_Li_Wan_2008}
Xie, F., Li, C., and Wan, G., ``An efficient and simple method of mpsk
  modulation classification,'' in [{\em 2008 4th International Conference on
  Wireless Communications, Networking and Mobile
  Computing}{\nolinebreak\hspace{0.1em}]},   1–3 (Oct 2008).

\bibitem{zhou2013design}
Zhou, P., An, Q., Xia, W., and He, Z.~S., ``Design and implementation of
  modulation recognition algorithm based on monitoring receiver,'' in [{\em
  Applied Mechanics and Materials}{\nolinebreak\hspace{0.1em}]},   {\bf 411},
  898--902, Trans Tech Publ (2013).

\bibitem{Zhang_Wang_Wu_Tang_2018}
Zhang, Y., Wang, J., Wu, G., and Tang, Q., ``Wireless signal classification
  based on high-order cumulants and machine learning,'' in [{\em 2018 IEEE
  International Conference of Safety Produce Informatization
  (IICSPI)}{\nolinebreak\hspace{0.1em}]},   246–250 (Dec 2018).

\bibitem{Rajendran_Meert_Giustiniano_Lenders_Pollin_2018}
Rajendran, S., Meert, W., Giustiniano, D., Lenders, V., and Pollin, S., ``Deep
  learning models for wireless signal classification with distributed low-cost
  spectrum sensors,'' {\em IEEE Transactions on Cognitive Communications and
  Networking}~{\bf 4},  433–445 (Sep 2018).

\bibitem{Guo_Jiang_Wu_Zhou_2020}
Guo, Y., Jiang, H., Wu, J., and Zhou, J., ``Open set modulation recognition
  based on dual-channel lstm model,'' {\em arXiv:2002.12037 [eess]}  (Feb
  2020).
\newblock arXiv: 2002.12037.

\bibitem{Hochreiter_Schmidhuber_1997}
Hochreiter, S. and Schmidhuber, J., ``Long short-term memory,'' {\em Neural
  Computation}~{\bf 9},  1735–1780 (Nov 1997).

\bibitem{OShea_Corgan_Clancy_2016}
O’Shea, T.~J., Corgan, J., and Clancy, T.~C., ``Convolutional radio
  modulation recognition networks,'' in [{\em Engineering Applications of
  Neural Networks}{\nolinebreak\hspace{0.1em}]},  Jayne, C. and Iliadis, L.,
  eds., {\em Communications in Computer and Information Science},  213–226,
  Springer International Publishing (2016).

\bibitem{West_OShea_2017}
West, N.~E. and O’Shea, T., ``Deep architectures for modulation
  recognition,'' in [{\em 2017 IEEE International Symposium on Dynamic Spectrum
  Access Networks (DySPAN)}{\nolinebreak\hspace{0.1em}]},   1–6 (Mar 2017).

\bibitem{Krizhevsky_Sutskever_Hinton_2017}
Krizhevsky, A., Sutskever, I., and Hinton, G.~E., ``Imagenet classification
  with deep convolutional neural networks,'' {\em Communications of the
  ACM}~{\bf 60},  84–90 (May 2017).

\bibitem{OShea_Roy_Clancy_2018}
O’Shea, T.~J., Roy, T., and Clancy, T.~C., ``Over the air deep learning based
  radio signal classification,'' {\em IEEE Journal of Selected Topics in Signal
  Processing}~{\bf 12},  168–179 (Feb 2018).
\newblock arXiv: 1712.04578.

\bibitem{Szegedy_Ioffe_Vanhoucke_Alemi_2016}
Szegedy, C., Ioffe, S., Vanhoucke, V., and Alemi, A., ``Inception-v4,
  inception-resnet and the impact of residual connections on learning,'' {\em
  arXiv:1602.07261 [cs]}  (Aug 2016).
\newblock arXiv: 1602.07261.

\bibitem{Cao_Xiao_Yang_Fang_Yang_Liu_Li_2019}
Cao, Y., Xiao, C., Yang, D., Fang, J., Yang, R., Liu, M., and Li, B.,
  ``Adversarial objects against lidar-based autonomous driving systems,'' {\em
  arXiv:1907.05418 [cs, stat]}  (Jul 2019).
\newblock arXiv: 1907.05418.

\bibitem{engstrom2019adversarial}
Engstrom, L., Ilyas, A., Santurkar, S., Tsipras, D., Tran, B., and Madry, A.,
  ``Adversarial robustness as a prior for learned representations,'' {\em arXiv
  preprint arXiv:1906.00945}  (2019).

\bibitem{ilyas2019adversarial}
Ilyas, A., Santurkar, S., Tsipras, D., Engstrom, L., Tran, B., and Madry, A.,
  ``Adversarial examples are not bugs, they are features,'' in [{\em
  NeurIPS}{\nolinebreak\hspace{0.1em}]},   125--136 (2019).

\bibitem{Goodfellow_Shlens_Szegedy_2015}
Goodfellow, I.~J., Shlens, J., and Szegedy, C., ``Explaining and harnessing
  adversarial examples,'' {\em arXiv:1412.6572 [cs, stat]}  (Mar 2015).
\newblock arXiv: 1412.6572.

\bibitem{Madry_Makelov_Schmidt_Tsipras_Vladu_2019}
Madry, A., Makelov, A., Schmidt, L., Tsipras, D., and Vladu, A., ``Towards deep
  learning models resistant to adversarial attacks,'' {\em arXiv:1706.06083
  [cs, stat]}  (Sep 2019).
\newblock arXiv: 1706.06083.

\bibitem{Kurakin_Goodfellow_Bengio_2017}
Kurakin, A., Goodfellow, I., and Bengio, S., ``Adversarial examples in the
  physical world,'' {\em arXiv:1607.02533 [cs, stat]}  (Feb 2017).
\newblock arXiv: 1607.02533.

\bibitem{dong2018boosting}
Dong, Y., Liao, F., Pang, T., Su, H., Zhu, J., Hu, X., and Li, J., ``Boosting
  adversarial attacks with momentum,'' in [{\em Proceedings of the IEEE
  conference on computer vision and pattern
  recognition}{\nolinebreak\hspace{0.1em}]},   9185--9193 (2018).

\bibitem{OShea_West_2016}
O’Shea, T.~J. and West, N., ``Radio machine learning dataset generation with
  gnu radio,'' {\em Proceedings of the GNU Radio Conference}~{\bf 1} (Sep
  2016).

\bibitem{Ilyas_Santurkar_Tsipras_Engstrom_Tran_Madry_2019}
Ilyas, A., Santurkar, S., Tsipras, D., Engstrom, L., Tran, B., and Madry, A.,
  [{\em Adversarial Examples Are Not Bugs, They Are
  Features}{\nolinebreak\hspace{0.1em}]},  125–136, Curran Associates, Inc.
  (2019).

\bibitem{Engstrom_Ilyas_Santurkar_Tsipras_Tran_Madry_2019}
Engstrom, L., Ilyas, A., Santurkar, S., Tsipras, D., Tran, B., and Madry, A.,
  ``Adversarial robustness as a prior for learned representations,'' {\em
  arXiv:1906.00945 [cs, stat]}  (Sep 2019).
\newblock arXiv: 1906.00945.

\bibitem{Ortiz-Jimenez_Modas_Moosavi-Dezfooli_Frossard_2020}
Ortiz-Jimenez, G., Modas, A., Moosavi-Dezfooli, S.-M., and Frossard, P., ``Hold
  me tight! influence of discriminative features on deep network boundaries,''
  {\em arXiv:2002.06349 [cs, stat]}  (Feb 2020).
\newblock arXiv: 2002.06349.

\bibitem{andrew1996tanenbaum}
Andrew, S., ``Tanenbaum computer networks,'' {\em Computer Networks, Englewood
  Cliffs} ,  141--148 (1996).

\end{thebibliography}
\bibliographystyle{spiebib} % makes bibtex use spiebib.bst

\end{document}